# Discrete Army Ant Search Optimizer-Based Target Coverage Enhancement in Directional Sensor Networks

Yin-Di Yao[1], Qin Wen[1*], Yan-Peng Cui[2*], and Bo-Zhan Zhao[1]

[1] School of Communication and Information Engineering and the School of Artificial Intelligence, Xi'an University of Posts and Telecommunications, Xi'an 710121, China
[2] The Key Laboratory of Universal Communications, Ministry of Education, Beijing University of Posts and Telecommunications, Beijing 100876, China
*Student Member, IEEE

*Abstract*—Coverage of interest points is one of the most critical issues in directional sensor networks. However, considering the remote or inhospitable environment and the limitation of the perspective of directional sensors, it is easy to form perception blind after random deployment. The intension of our research is to deal with the bound-constrained optimization problem of maximizing the coverage of target points. A coverage enhancement strategy based on a discrete army ant search optimizer (DAASO) is proposed to solve the above problem, which is inspired by the biological habits of army ants. A set of experiments are conducted using different sensor parameters. Experimental results verify the effectiveness of the DAASO in coverage effect when compared to the existing methods.

## I. INTRODUCTION

The wireless sensor network (WSN) paradigm sees smart machines capable of accessing data about targets without human interaction [1]. As an important part of WSNs, a directional sensor network (DSN) has been the focus of research community with the application of directional sensors (i.e., video sensor and camera) [2]. Coverage is a basic application of DSNs, which aims to gain enhanced observations in the field of interest [3]. Considering the limitation of the viewing angle of directional sensors and the harsh monitoring environment, it is difficult to guarantee the quality of coverage [4]. Therefore, how to adjust the sensing direction of nodes to achieve coverage enhancement is an urgent problem of DSNs.

As the most representative and effective research, Zhu *et al.* [5] proposed three heuristics to solve the problem of deploying environmental energy harvesting DSNs with the minimum cost, where targets have different coverage requirements. Zhu *et al.* [6] formulated and addressed the minimum-cost DSN deployment problem by using greedy heuristic, local search, and particle swarm optimization (PSO). Ai and Abouzeid [7] presented a centralized greedy algorithm to solve the maximum coverage problem with the fewest sensors of DSNs. In [8], PSO is combined with a boundary constraint of eliminating the redundancy scheme to satisfy the demand of the target area. Utilizing improved adaptive particle swarm optimization (IAPSO), the authors achieved the high coverage rate and the low redundancy ratio [9]. As one of the most classical evolutionary algorithms, the genetic algorithm (GA) and its derivative algorithms are usually utilized for solving the coverage problem. In [10], a *K*-coverage model based on the GA is proposed to achieve continuous coverage and extend the lifetime.

Corresponding author: Qin Wen (e-mail: wq199802@163.com).
Associate Editor: F. Falcone.
This work was supported in part by the Agricultural Project of Science and Technology Department of Shaanxi Province under Grant 2021NY-180, and in part by the Shaanxi Province Innovative Talents Promotion Plan—Internet of Things Technology Innovation Team under Grant 2019TD-028.
Digital Object Identifier 10.1109/LSENS.2022.3158274

Fan *et al.* [11] obtained the maximum effective coverage rate of DSNs by introducing the quantum GA. The efficiency of metaheuristic in solving the coverage of WSNs has been proved by many scholars.

In this letter, we address a practically important problem of maximizing the coverage of target points by adjusting the sensing directions of directional sensors, which is an NP-hard problem that can be solved by using the swarm intelligence (SI) algorithm [7]. The No Free Lunch theorem has logically proved that there is no optimization algorithm best suited for solving all the optimization problems [12]. The existing SI algorithm is easy to fall into the local optimal solution when solving the coverage problem of DSNs, which results in an upper limit of coverage. The intention of our research is to propose a novel SI algorithm, which can effectively solve the target coverage problem of DSNs when compared with the existing SI algorithms. The major contributions of our research are as follows:
1) A discrete army ant search optimizer (DAASO) is first proposed inspired by the predatory behavior of the army ant.
2) The DAASO is utilized for maximizing the coverage of target by integrating the corresponding relationship between the DAASO and the DAASO-based target coverage enhancement in DSNs.

## II. SYSTEM MODEL

We consider DSNs of *D* directional sensor nodes and *M* targets in the monitoring area $L \times W$. As shown in Fig. 1(a), there is a control center (e.g., a sink), which collects location information and broadcasts their orders to sensors. We choose not to discuss the positioning and route problems in this letter due to the space limitation and pay more attention to the target coverage problem.

Different from the existing isotropic sensor, the sensing region of a directional sensor is limited by the viewing angle, as shown in Fig. 1(b). The directional sensing model can be expressed as a seven-tuple $< (x_i, y_i), R, \vec{V}, \theta, p, \alpha >$, where $(x_i, y_i)$ is the position of the sensor, $R$ is the sensing radius, $\vec{V}$ is the sensing direction, and $\theta$ is the angle of view. Additionally, each of sensor has *p* possible sensing directions, a sensor can only choose one active sector at any time, and $\alpha$ is the deviation angle between $\vec{V}$ and $x$-axis.

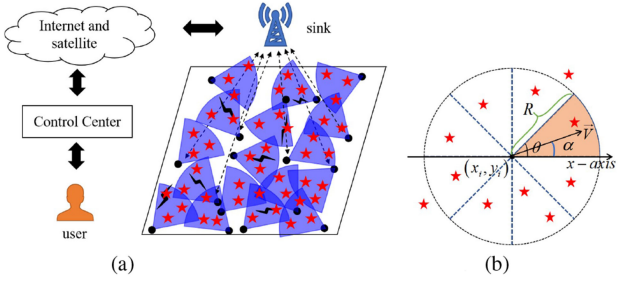

Fig. 1. Coverage and directional sensing models. (a) Coverage model. (b) Sensing model.

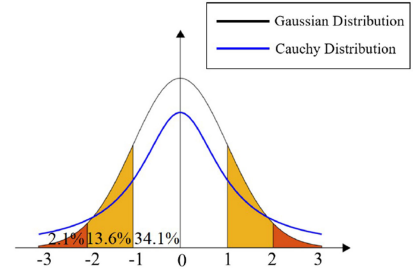

Fig. 2. Gaussian distribution and Cauchy distribution.

As the aforementioned system model shows, not all the targets are covered under random deployment; our intention is to change the initial orientations and cover as many targets as possible, with the constraint that only one orientation can be chosen for a sensor. We call this the maximum target coverage (MTC) problem.

The MTC problem can be stated as

$$\begin{aligned} \max \quad & \sum_{k=1}^{M} \psi_k \\ \text{s.t.} \quad & \sum_{j=1}^{p} \chi_{i,j} = 1 \quad \forall\, i = 1, \ldots, D \\ & \psi_k = 0 \text{ or } 1 \quad \forall\, k = 1, \ldots, M \end{aligned} \quad (1)$$

where $D$ is the number of sensors, $M$ is the number of targets, $\psi_k$ is a binary variable that has value 1 if target $k$ is covered by any sensor, and 0 otherwise, and $\chi_{i,j}$ is a binary variable that has value 1 if the directional sensor $S_i$ uses the sensing direction $j$.

## III. PROPOSED ALGORITHM

In order to solve the MTC problem, we propose the discrete target coverage enhancement strategy based on the DAASO according to the biological habits of the army ant. In this section, we provide the mathematical models of predation behavior, following, attacking, and discretization. Then, the DAASO-based target coverage enhancement in DSNs is outlined.

### A. Predation Behavior

Army ants are dominant social hunters of other ants, mainly living in the Amazon River Basin. There exist clear mechanisms of choosing the prey in army ant population. Specifically, the army ants use specific prey odors, including alarm odors, dead ants, live ants, and nest material, to detect potential prey and direct their foraging. Additionally, they have the strongest response to the nest material of their preferred prey, with progressively weaker responses across the live ant, dead ant, and alarm odor treatments [13]. In order to mathematically model the mechanism of choosing the prey, we consider the best fitness solution as the nest material; the next three solutions are named live ant, dead ant, and alarm odors.

In the proposed DAASO, it is assumed that the problem's variables are the position of army ants in the space; the army ants can move in 1-D, 2-D, or hyperdimensional space with changing their position vectors. Therefore, the position of army ant population can be expressed as $Ant^{N \times D} = [Ant_{i,1}, Ant_{i,2}, \ldots, Ant_{i,D}]$, $i = 1, 2, \ldots, N$, where $N$ is the number of army ants and $D$ is the number of variables (dimension).

### B. Following

Army ants follow interesting odors to move during the forage. The following behavior is mathematically modeled as the following equation:

$$\text{Ant}_{i-j}^{1 \times D}(t+1) = \text{Prey}_j^{1 \times D}(t) + \left(\text{Prey}_j^{1 \times D}(t) - \text{Ant}_i^{1 \times D}(t)\right) \Theta N(0, 1)^{1 \times D} \quad (2)$$

where $t$ is the current, iteration, and $\text{Ant}_i^{1 \times D}$ and $\text{Prey}_j^{1 \times D}$ indicate the position of army ant $i$ and the position of prey $j$, respectively. $\Theta$ is the Hadamard product operator and $N(0, 1)^{1 \times D}$ is a vector of $1 \times D$ dimension, in which the elements are random numbers with standard normal distribution. The propose of this design is to make a good compromise between exploration and exploitation in the DAASO. Given a random number $\epsilon$ following standard Gaussian distribution, as shown in Fig. 2, the probability of $|\varepsilon| < 1$ is 68.2%, which reflects the exploitation of the DAASO. On the contrary, it reflects the exploration of the DAASO. The design of $|\varepsilon| > 1$ also considers the influence of obstacles to army ants during hunting path, which forces the army ants to expand their search space to find a better solution.

### C. Attacking

Army ants have the ability to differentiate and use the location of prey odors. The attack direction and position are usually guided by the prey odors. The following formulas are proposed to describe this process:

$$\text{Ant}_i^{1 \times D}(t+1) = \sum_{j=1}^{\text{num}_{\text{ant}_i}(t)} \text{Ant}_{i-j}^{1 \times D}(t+1) / \text{num}_{\text{ant}_i}(t) \quad (3)$$

where $\text{num}_{\text{ant}_i}(t)$ is the number of odors that attract $\text{Ant}_i$ in the $t$th iteration. It is worth noting that not every army ant can smell the four odors due to the influence of weather, distance, odor volatilization, and other factors. The number of army ants attracted by odors in each iteration is described by Poisson distribution, which is suitable for expressing the number of random events per unit time. Specifically, assuming that the number of army ants attracted by odors in each iteration follows Poisson distribution, that is, the probability that there are $k$ army ants in the population $Ant$ attracted by odors in the $t$th iteration is defined as

$$P(k \text{ army ants in } \mathbf{Ant}) = \frac{(\text{num}_{\text{ave}}(t))^k}{k!} e^{-\text{num}_{\text{ave}}(t)}, \\ k = 0, 1, 2, \ldots, N \quad (4)$$

where $\text{num}_{\text{ave}}(t)$ is the average value of army ants attracted by odors in the $t$th iteration. In order to increase the convergence of the DAASO, the number of army ants attracted by odor increases with the increase

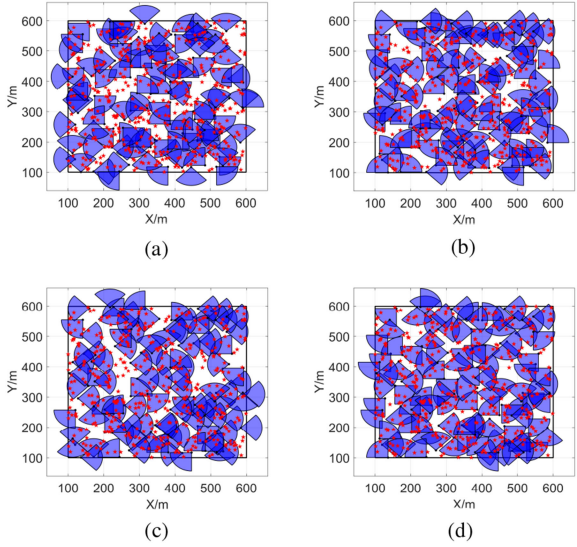

Fig. 3. Comparison of coverage effect. (a) Initial coverage. (b) IAPSO. (c) DGWO. (d) DAASO.

in the number of iterations

$$\text{num}_{\text{ave}}(t) = \text{num}_{\text{ini}} + (N - \text{num}_{\text{ini}}) \times \frac{t}{T_{\max}} \quad (5)$$

where $\text{num}_{\text{ini}}$ denotes the average number of army ants initially recruited by prey, and its value is $N/2$; $T_{\max}$ is the maximum number of iterations. The exact number of army ants recruited by odors can be selected by roulette, and the index of army ant can be randomly selected from the whole population.

Army ants will randomly follow the two companions in the population to move once they are not attracted by any kind of odor

$$\text{Ant}_i^{1 \times D}(t+1) = \sum_{k=1}^{2} \left(\text{Ant}_k^{1 \times D}(t) + \text{Cauchy}(0,1)\right)^{1 \times D} / 2 \quad (6)$$

where $\text{Cauchy}(0,1)^{1 \times D}$ is a vector of $1 \times D$ dimension, in which the elements are random number with standard Cauchy distribution. As shown in Fig. 3, the Cauchy distribution has better perturbation compared with the Gaussian distribution, which help the DAASO explore more local space.

Additionally, with the raids of army ants, the number of preys continues to decline. The mathematical modeling is

$$\text{preynumber}(t) = \text{round}\left(4 - 4 \times \frac{t-1}{T_{\max}}\right). \quad (7)$$

### D. Discretization

The army ant searches prey based on continuous space. However, the MTC problem of DSNs, as shown in (2), is the discrete optimization problem. Therefore, the data discretization is performed after the position of the army ant is obtained in each iteration. Specifically, we assume that there are $M$ discrete variables $(V_1, V_2, \ldots, V_M)$, and the discrete interval between adjacent variables is $\alpha$. For any value $X \in (V_i, V_{i+1})$, the discretization operation of $X$ is

$$X_{\text{new}} = V_i + \alpha \times b \quad (8)$$

$$b = \begin{cases} 1, & r_2 < \frac{X - V_i}{V_{i+1} - V_i} \\ 0, & r_2 > \frac{X - V_i}{V_{i+1} - V_i} \end{cases} \quad (9)$$

Table 1. Corresponding Relationship Between the Daaso and the Target Coverage Based on the DAASO

| DAASO | Target Coverage Based on DAASO |
|---|---|
| Population size | Number of deployment schemes |
| Number of variables (dimension) | Number of directional sensors |
| Position of army ant | Sensing direction |
| Fitness | Number of coverage targets |
| Best fitness | Maximum number of coverage targets |
| Number of discrete variables | Number of possible sensing directions |

**Algorithm 1:** DAASO-Based Target Coverage Enhancement in DSNs.

**Require:** Sensor nodes: $S = \{S_1, S_2, \ldots, S_D\}$; Predefined army ant size: $N$; Number of dimensions of army ants: $\text{num} = D$; Possible sensing directions: $p$; Maximum iterations: $T_{\max}$.
**Ensure:** The sensing direction of sensors; maximum number of coverage targets.
1: Initialize army ants (sensing directions): $\vec{V}_{i,j}, \forall i, j, 1 \leq i \leq N, 1 \leq j \leq D$
2: Discrete the position of army ant by Section III-D.
3: **for** $i = 1 \rightarrow N$ **do**
4:    Calculate the fitness (number of coverage targets) of $\vec{V}_i$.
5: **end for**
6: Generate initial position of prey.
7: **for** $t = 1 \rightarrow T_{\max}$ **do**
8:    **for** $i = 1 \rightarrow N$ **do**
9:      Update the position of $\vec{V}_i$ by Sections III-B and III-C.
10:    **end for**
11:    Discrete the position of army ant $\vec{V}$ by Section III-D.
12:    Calculate the fitness of $\vec{V}$.
13:    Update the number, position, and fitness of prey.
14: **end for**
15: Acquire the optimal sensing direction $\vec{V}_i *$ of sensors and the maximum number of coverage targets.

where $r_2$ is a random number $(0, 1]$; it can be intuitively seen that the setting of $b$ provides the ability of mutation for the DAASO.

### E. DAASO-Based Target Coverage Enhancement

The corresponding relationship between the DAASO and the target coverage enhancement based on the DAASO is shown in Table 1. Algorithm 1 shows the pseudocode of the DAASO-based target coverage enhancement in DSNs. The main steps are as follows.

*Step 1:* Initialize the directional sensor $S = \{S_1, S_2, \ldots, S_D\}$ and army ant $\vec{V}_i = [\vec{V}_{i,1}, \vec{V}_{i,2}, \ldots, \vec{V}_{i,k}, \ldots, \vec{V}_{i,D}], i = 1, 2, \ldots, N$; each army ant represents a deployment scheme containing $D$ sensing directions. For example, $\vec{V}_{i,k}$ is the sensing direction of sensor $S_k$ in the $i$th deployment.

*Step 2:* Discrete the position of each army ant and calculate the number of coverage targets (NCT) by each army ant. The best four fitness obtained by initial positions of army ants saved as initial prey.

*Step 3:* Update and discrete the position of each army ant according to the following operator, the attacking operator, and the discretization operator.

*Step 4:* Calculate the NCT by each army ant and update the fitness, position, and number of prey.

*Step 5:* If the maximum number of iterations is exceeded, the optimal sensing direction is obtained and the algorithm ends. Otherwise, proceed to Step 3.

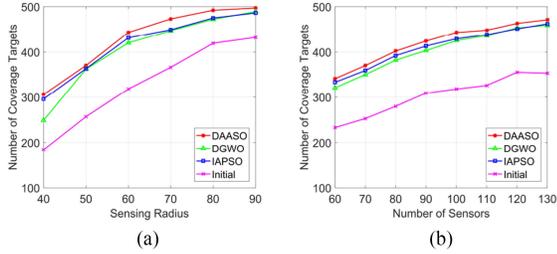

Fig. 4. (a) and (b) Relationship between the NCT with the sensing radius and the number of multimedia sensor nodes, respectively.

Table 2. Performance Comparison of the Average Results of 50 Experiments

| Parameter | DAASO | DGWO | IAPSO |
|---|---|---|---|
| $M = 500, R = 60m, D = 100, \theta = \pi/2$ | 443 | 421 | 433 |
| $M = 500, R = 40m, D = 100, \theta = \pi/2$ | 306 | 289 | 296 |
| $M = 500, R = 60m, D = 100, \theta = \pi/3$ | 380 | 364 | 362 |
| $M = 500, R = 60m, D = 60, \theta = \pi/2$ | 341 | 334 | 332 |

## IV. EXPERIMENTAL RESULTS AND DISCUSSION

In this section, a set of experiments are performed with MATLAB 2016a to evaluate the performance of our proposed algorithm. By taking the NCT as the fitness value, we compare the performance of DAASO, IAPSO [9], and gray wolf optimizer (DGWO) [14] about the NCT in DSNs under different scenes, where the relevant parameters in IAPSO and DGWO adopt the values set in the original algorithm. The population size and the maximum number of iterations are set to 50 and 100, respectively.

When the number of targets is 500, a total of 100 sensors ($R = 60$ m, $\theta = \pi/2$, $\alpha = \pi P/4$, $P = 1, 2, \ldots, 8$) are deployed in the monitoring area of 500 m $\times$ 500 m. Fig. 3 shows the initial coverage effect and the final coverage effect of the considered three algorithms. It is intuitive to see that the DAASO shows better performance in terms of the NCT. The initial NCT is 327; specifically, the final NCT of IAPSO and DGWO are 443 and 430, respectively. The final NCT of the DAASO can reach up to 459. Fig. 4 shows the plots of the NCT with the sensing radius and the number of sensors. Intuitively, the NCT enhances with the increase in the sensing radius and the number of sensors. Additionally, the NCT of the DAASO is always higher than those of other two algorithms.

In order to verify the reliability of the DAASO for the NCT, 50 experiments are performed independently with different scenes, as shown in Table 2. Intuitively, the simulation results can firmly prove the reliability of the DAASO and verify the accuracy of the previous single experiment. The reasons leading to the performance differences of the final NCT of DAASO, DGWO, and IAPSO are analyzed in detail as follows. DAASO, DGWO, and IAPSO all initiate optimization process by the movement of search agents in the search space; however, the movement mechanism is entirely different. In IAPSO, new step and direction of movement are obtained by velocity, *pbest*, and *gbest*; it is noteworthy that the rotation angle of the sensors so as to the final NCT will be inevitably affected by the threshold of the velocity and the weight. The adaptive weight in IAPSO provides more exploration when compared with classical PSO. However, it is difficult to set a suitable velocity threshold to ensure NCT, namely, the performance of the final NCT is still worse than that of the DAASO. In essence, both the DAASO and the DGWO use better agents to guide the position update of particles. However, technically, both present several differences in their formulation and updating mechanism. In the DGWO, new direction for the movement of the gray wolf agent is obtained by alpha wolf, beta wolf, and omega wolf, namely, the cumulative effect of the three is considered. In the DAASO, odor sources are considered as solutions for an optimization problem. However, the cooperation of multiple operators in the DAASO effectively balances exploration and exploitation compared with the DGWO. Specifically, a new solution is generated by following, attacking, and discretization operators. In the following operator, the army ant will randomly appear around the odor of prey with normal distribution, which provides the DAASO with the global exploration. In other words, it forces army ants apart from the prey and explores more solutions. In the attacking operator, the design of Cauchy distribution ensures the diversity of population. In the target coverage problem of DSNs, it can effectively help the DAASO to explore more deployment schemes. In addition, a discrete operator essentially provides variability for the DAASO, which allows a single sensor to find more uncovered targets in the neighboring space.

## V. CONCLUSION

In this letter, the optimization problem about the maximization of interesting targets in DSNs was solved efficaciously by the proposed strategy DAASO, which has superior performance in terms of NCT when compared with IAPSO and DGWO. We firmly believe that the DAASO will perform better in other optimization, for example, how to use the DAASO to solve the minimum-cost deployment problem [6] will be the focus of the next research.